\documentstyle[12pt]{article}

\begin{document}
\begin{titlepage}
\begin{centering}
{\Huge\bf Ordered Phase and\\ Field-Induced Domains in
a\vspace{0.8cm}\\ Short-Range Ising Spin Glass}\\
\vspace{1.5cm}
{\Large E.P. Raposo$^\dagger $ \Large and
M.D. Coutinho-Filho$^\ddagger $}\\
\vspace{1.1cm}
{\large Depto. de F\'{\i}sica, Universidade Federal de Pernambuco\\
Cidade Universit\'aria, 50670-901, Recife, PE, Brazil
\vspace{0.2cm}\\
\hspace{4.0cm}
$\dagger $ ecpr@npd1.ufpe.br, $\ddagger $ 10mdcf@npd.ufpe.br
}
\end{centering}
\vspace{0.7cm}
\large
\begin{abstract}
\large
\noindent
Using a microscopic numerical approach suitable to describe disordered
 antiferromagnets, with application to $Fe_{x}Zn_{1-x}F_{2}$, it is
 shown that the characteristics of the spin glass phase found for
 $x=0.25$ is much in agreement with the scenario predicted by the
 scaling theory of the droplet model
.\\\\
\normalsize
%\noindent
\end{abstract}
\end{titlepage}
\def\carre{\vbox{\hrule\hbox{\vrule\kern 3pt
\vbox{\kern 3pt\kern 3pt}\kern 3pt\vrule}\hrule}}
\large
\newpage
%%%\begin{texto}

The Edwards-Anderson (EA) model[$1$] has become the prototype
 model used to perform most
theoretical studies of spin glasses (SGs). The central
 feature here is to capture the two essential ingredients,
 namely disorder and frustration. The latter is introduced
 by competing ferromagnetic (F) and antiferromagnetic (AF)
 interactions, while the former requirement is built into
 the model by allowing the exchange couplings to have a given
 probability distribution. A simple solution[$2$] of the
 model was attempted by assuming infinite-range exchange
 interactions (SK model), but it was later shown[$3,4$]
 to be a quite hard and complex task. This rather unusual
 mean-field theory is characterized by ultrametric-based
 hierarchy, with infinite degenerate ground states
 unrelated by any obvious (if any) symmetry.
 This picture proved to be of
undoubted importance in several fields such as biology and
 optimization problems[$5$], where large effective
 coordination numbers mimic a high-dimensional space.
 The question we address, however, is that related to the
 original motivation: is the mean-field theory  a good
 starting point for real short-range (Ising for simplicity)
 three-dimensional SGs?
This question gave rise to several controversies as a more
 phenomenological approach, namely the ``droplet model''[$6,7$]
, offered a quite distinct alternative scenario. The droplet
 model is based on numerical results of the short-range EA
 model, supplemented by scaling and renormalization-group
 ideas. In particular, it predicts a nondegenerate ground
 state, except for the trivial up and down possibilities.
 Moreover, in the droplet picture the SG phase is destroyed
 by an applied external magnetic field ($H$) and thus,
 contrary to the mean-field prediction, no {\it equilibrium}
 de Almeida-Thouless (AT) line exists.  Experimental[$8$]
 and theoretical studies[$9$] facing these controversies have
 found it difficult to announce clear-cut results,  in part
 because of the very slow dynamics plaguing these systems,
 a characteristic shared by both competitive descriptions.\\
In this work we study the above-mentioned problems using a
direct microscopic approach suitable to describe short-range
 three-dimensional insulating Ising SGs. A random-site
 dilution is thus used instead of the random-bond approach of
 the EA model. In fact, we shall parametrize our model
 Hamiltonian in the expectation that a description of the
 experimental observations[$10$] on the compound
 $Fe_{x}Zn_{1-x}F_{2}$, as a function of dilution $x$, is
achieved.\\
Very recently, we  showed[$11$] that the presence of a very
 small frustrated interaction is the mechanism underlying the
 appearence of the SG phase at $H=0$ in
 $Fe_{0.25}Zn_{0.75}F_{2}$. Here we extend our study and calculate
 the spin-spin correlation function in the ordered SG phase
 and  the characteristics of  the field-induced domains both
in the SG and in the ``random field Ising model" (RFIM) phases.
 While the RFIM phase is very stable under an applied field,
 its effect on the SG phase is dramatic regardless of the value
 of the field. Our results are in agreement with the scenario
 predicted by the droplet model, including the chaotic behavior
 of the correlation function as the spin distance varies at
 constant $T$. The main quantitative result is the value of
 the exponent governing the $H = 0$ SG ordered phase,
 $y_{T} = 0.19 \pm 0.03$. Moreover, we show that an
 irreversibility (AT) line exists regardless of the presence
 of frustration. We conclude that these lines are metastable
 lines, as also supported by the droplet model, and that the
 SG phase is indeed destroyed by an applied field, as recently
 experimentally reported for the compound
 $Fe_{0.5}Mn_{0.5}TiO_{3}$ [$12$].\\
{\it Formal procedure}: we perform numerical simulations of
the iterative set of equations of the local (site-by-site)
 mean fields (LMFs) derived from the microscopic Hamiltonian
\begin{equation}
{\cal H} = \sum_{<i,\delta_{\ell}>}J_{\ell}{\cal E}_{i}{\cal E}_{i+
\delta_{\ell}}S_{i}S_{i+\delta_{\ell}} -
\mu_{o} \sum_{i} {\cal E}_{i}S_{i}H,
\end{equation}
where $S_{i} = \pm 2$ (as in $Fe^{+2}$), ${\cal E}_{i} = 0, 1$
 and ${\ell}$ is summed over the three nearest-neighbor exchange
 interactions $J_{\ell}$ of the centred tetragonal lattice of
 $Fe_{x}Zn_{1-x}F_{2}$ [$10$]. We choose the values of the
 exchange constants by keeping unaltered the experimental
ratios $j_{1} = J_{1}/J_{2} =
- 0.013 , j_{3} = J_{3}/J_{2} = + 0.053$  and such as to
 fix the N\'eel temperature of the pure system,
$T_{N}(x=1) = 77.8 K$.
 This LMF technique has proven[$13$] very efficient in describing
 inhomogeneous and disordered systems. The thermally
averaged local spin $m_{i}$, results from the free energy
minimization in the form
 (${\cal E}_{i} = 1$):
\begin{equation}
m_{i} = < S_{i} >_{T} =  2\tanh \{ 2 h_{i}/KT \}  ,
\end{equation}
where the local field at site $i$ is given by
\begin{equation}
h_{i} = -\frac{1}{2} \sum_{\delta_{\ell}} J_{i+\delta_{\ell}}{\cal E}_{i+
\delta_{\ell}}m_{i+\delta_{\ell}} + \mu_{o} H  .
\end{equation}
Introducing $C_{ij}(T) = <S_{i}S_{j}>_{T} - m_{i} m_{j}$, we
find the LMF expression
\begin{equation}
<S_{i}S_{j}> = 4 \frac{\cosh [2(h_{i} + h_{j})/KT] - \cosh [2(h_{i} - h_{j})
/KT]}{\cosh [2(h_{i} + h_{j})/KT] + \cosh [2(h_{i} - h_{j})/KT]} .
\end{equation}
The spin-spin  correlation function of interest is then
$G_{T}(r_{ij}) = $
\linebreak
$<C_{ij}^{2}>_{av}$ ,
where $av$ indicates an average over distinct random spin
 configurations, such that
$<C_{ij}>_{av} = 0$.
The staggered magnetization is defined by
\begin{equation}
M_{S}(T,x) = (2/N)\sum_{i}{\cal E}_{i}m_{i} ,
\end{equation}
with $i$ summed over a given sublattice, and the SG-EA order
parameter by
\begin{equation}
Q(T,x) = (1/N)\sum_{j}{\cal E}_{j}<S_{j}^{2}>_{T},
\end{equation}
with $j$ taken through the whole lattice.\\
We choose a random initial configuration $\{S_{i},{\cal E}_{i}\}$
 in the high-$T$ paramagnetic phase and average over $50$
 independent runs. The system is cooled by an amount $\Delta T$
 ( we take $\Delta T = 0.05 K$ for the (H,T) phase diagram ,
 and $\Delta T = 1 K$ otherwise ), using ($2$) and a proper
 convergence criterion[$11,13$], down to $T = 2 K$ and heated back
 using the same ammount of $\Delta T$. We take $N = 2\cdot 30^{3}$
 sites and use periodic boundary conditions.\\
{\it Numerical data and analysis}: first we point out that our
 data correspond to a proper time scale in which a quasi-equilibrium
 state is achieved at $H = 0$. In Fig.$1(a)$ we illustrate a SG
 orderered phase configuration through  part of a randomly chosen
 sheet: $x= 0.25$,  $T/T_{N}(x=1) = 0.18$ and $H = 0$. The local-field
 distribution corresponding to this state is shown in Fig.$1(b)$.
 As evidenced  from the inset, competitive local fields exist only
 in the presence of frustration. In fact, the effect of frustration
 manifests itself only in the high dilution regime ( $x < 0.5$ )
 and causes domains of reversed spins surrounded by domain walls
 already when the system behaves as a RFIM ( $x = 0.48$ ) [$11$].
 This spin reversal effect  completely changes the balance of forces,
 thus causing the appearence of  competitive local fields,
 particurlarly near the percolation threshold  in which case
 a full SG phase emerges.\\
In the scaling theory[$6,7$] of the droplet model the SG ordered
 phase is characterized by the exponent associated with the
 thermal eigenvalue , $(T(L)/J(L)) \sim L^{-y_{T}}$, of the $T = 0$
 fixed point, where $J(L)$ is an effective exchange interaction.
 It also measures the scale of the free energy, $F = Y(T) L^{y_{T}}$,
  associated with the thermally activated droplets of reversed spins
 surrounded by domain walls. The average size of the domains increases
 as the temperature decreases, with $Y(T = 0) \sim J$. We have
 measured the correlation function $G_{r}$, for $x = 0.25$, and assumed
 a scaling dependence, $G_{r}(T) = f(r)\exp (-r/\xi )$, where
 $\xi \sim T^{(1/y_{T})}$, valid in the low-temperature region
 of the $H = 0$ SG ordered phase. The fitting of the data is shown
 in Fig.$2$: we find $y_{T}= 0.19 \pm 0.03$, in very good agreement
 with the estimates[$6$] using $T=0$ renormalization-group treatments
 of the short-range EA model. The data are for pairs of spins ( ten
 pairs in this case) inside the same domain. For pairs with spins in
 different domains the correlation is null.
Notice that the $T$-behavior
 of $G_{r}(T)$, shown in the  inset of Fig.$2$, confirms a finite-$T$
  phase transition. In Fig.$3(a)$ we plot the temperature dependence
of  $< S_{i} S_{j} >_{T}$ , with sites $i$ and $j$ randomly chosen on
 the lattice. Different behaviors are seen, including a change of
sign, indicating spin reversal by thermal activation of domains. On
 the other hand, the chaotic behavior of $< S_{i} S_{j} >_{T}$  as a
 function of spin distance at $T/T_{N}( x = 1) = 0.18$ is nicely
 displayed  in Fig.$3(b)$. The Lyapunov exponent associated with this
 bevavior is predicted[$7$] to be $\zeta = d_{s}/2 - y_{T}$, where
$d_{s}$ is the fractal dimension of the domain walls. $\zeta $ is
exactly one in $d = 1$ and expected to change very little with  the
 system dimensionality. Last, we should emphasize that below
$T/T_{N}( x = 1) = 0.18$ there are practically no spin reversal and
thus for all practical purposes the spin configuration of
Fig. 1(a) is the $H=0$ SG ground state {\it for a given
initial configuration}
$\{S_{i},{\cal E}_{i}\}$ in the paramagnetic phase;
$\{-S_{i},{\cal E}_{i}\}$ leads to a ground-state configuration with
 {\it all} spins reversed. Moreover, for a quenched disorder
configuration $\{{\cal E}_{i}\}$, the ground state remains the
{\it same} regardless of the spin paramagnetic configuration
$\{S_{i}\}$, i.e., the SG state is {\it self-averaged}[$9$].\\
 We now turn to study the system in the presence of an applied fied.
 In a field-cooled (FC) cycle random fields of strong magnitudes
induce SG-like states, {\it regardless} of the presence of frustration.
 These spin structures are formed by domains of the two (up and down)
 equivalent $H = 0$ SG-ground states[$6,7$], i.e., the original
 configuration of Fig.$1$(a) breaks up into domains of length
scale, $L_{H} \sim  (J/H)^{1/(d/2-y_{T})}$, which independently
align with the field. It means that the field is a relevant scaling
 variable governed by the (positive) exponent $y_{H} = d/2 - y_{T}$,
 $(H(L)/J(L)) \sim L^{y_{H}}$. The spin structures of these SG-like
states are illustrated in Fig. 4(a) by indicating the domains of
reversed spins ($\sqcap$) relative to the $H=0$ SG ground-state
(Fig. 1(a)), for $x = 0.25, T/ T_{N}(x=1) = 0.18$ and $H=1.0 G$.
The associated distribution of local fields are shown in Fig. 4(b).
 The inset of Fig. 4(b) shows that, in the absence of frustration,
 a zero-field-cooled (ZFC) cycle induces a metastable long-range
 highly nonuniform AF order as a result of the local AF interactions
predominating in the system. We should stress that we have access
 to the microscopic local spin values on our chosen sheets. It is
 thus observed that the local spin values on the
structure of Fig. 4(a)
 and on the mentioned AF structure (inset of Fig. 4(b)) are the same
of those in Fig. 1(a) ( for any $H < 10^{3} G$ ), except for a change
 of sign for those spins suffering a spin reversal and a small
magnetization component in the field direction ( experimentally
called[$10$] $M_{FC}$ and $M_{ZFC}$, respectively). This
is so because
 our fields states were ``prepared" using the {\it same} initial
configuration of Fig. 1(a) .\\
The characteristics of these field-induced states can also be studied by
 measuring the effect of frustration on the EA-SG order parameter.
In Fig.$5(a)$  we see the qualitative change of behavior as a result of
 an applied field in a FC cycle: while in the $H = 0$ state frustration
 is a fundamental ingredient ( $Q = 0$ if $j_{3} = 0$), under a field,
 regardless of its value, frustration  decreases the $Q$ value !
 Saturation of $Q$ is achieved at about $10^{2} T$. To further evidence
 the special character of the $H=0$ SG phase found for $x= 0.25$,
we have also studied the effect of frustration on the system
for $x = 0.48$,
 a regime dominated by random-field effects on top of the long-range AF
ordering. In Fig.$5(b)$ we see that frustration always decreases $Q$, even
  for $H = 0$. The effect of frustration peaks at an intermediate
 temperature, because in this situation  neither full ordering nor
 full disorder predominates in the system.\\
Finally, in Fig.$6$ we display the field-induced  AT - irreversibility
 lines ($\bullet $) obtained by the loci of the points in which $M_{FC}$
 differs from $M_{ZFC}$. We see that exponents compatible with the
 experimental results are found regardless of the presence
 of frustration.
 As we see from the insets of Fig. 6(a) and 6(b) the FC ($\sqcap$)
 and ZFC ($\diamond$) correlation functions, for $H = 0.5G$, peak at
 a temperature below the irreversibility lines, much in agreement
 with the $H=0$ correlation function in the presence of frustration.
 The data are for sites in the same domain. In the absence of
 frustration, however, the $H=0$ correlation function is null
 (inset of Fig. 6(a)). These results show that in our model for a
 highly disordered magnetic system, metastable ``glassy" states,
 characterized by the domains of up and down $H = 0$
 spin configurations
 of a SG state at a fixed temperature, are induced by an applied
 field as a result of the strong random fields generated in the
 system[$14$]. The effect of frustration is only relevant at very low
fields, in which case the approach to SG criticality is only felt
by the frustrated system. Though a specific low-temperature
 SG configuration
 depends on a given initial paramagnetic configuration
 $\{S_{i},{\cal E}_{i}\}$, the ordered phase is univocally
 characterized by the thermal eigenvalue $y_{T}$ of the $T=0$
 fixed point.\\
In conclusion, we have presented results based on a local mean-field
 numerical simulation of disordered AFs, with application to
 $Fe_{x}Zn_{1-x}F_{2}$, describing several features of these
systems as a function of concentration. In particular, for a
concentration near the percolation threshold, the system exhibits
 a SG phase with a scenario in agreement with the scaling theory
 of the ``droplet" model.\\
The authors acknowledge F.C. Montenegro, D.P. Belanger and J.R.L.
 de Almeida for several fruitful discussions and suggestions.
This work was supported by FINEP, CNPq, CAPES and FACEPE
 (Brazilian Agencies).\\
%%%\end{texto}
\newpage
%%%\begin{References}
{\bf References}\\\\
\begin{itemize}
\item{[1]} S. Edwards and P. Anderson, J. Phys. F\underline{5},
 965 (1975).
\item{[2]} D. Sherrington and S. Kirkpatrick, Phys. Rev. Lett.
 \underline{35}, 1792 (1975).
\item{[3]} J.R.L. de Almeida and D. Thouless, J. Phys.
 A\underline{11}, 983 (1978).
\item{[4]} G. Parisi, Phys. Rev. Lett. \underline{43}, 1754
 (1979); J. Phys. A\underline{13}, 1101 (1980); \underline{13},
1887 (1980); \underline{13}, L115 (1980); Phys. Rev. Lett
\underline{50}, 1946 (1983).
\item{[5]} M. M\'ezard, G. Parisi and M.A. Virasoro, {\it Spin
Glass Theory and Beyond} (World Scientific, Singapore, 1987).
\item{[6]} L. McMillan, Phys. Rev. B\underline{30}, 476 (1984); A.J.
 Bray and M.A. Moore, J. Phys. C\underline{17}, L463 (1984); L613
(1984); D.S. Fisher and D.A. Huse, Phys. Rev. Lett. \underline{56},
 1601 (1986); A.J. Bray, Comments Cond. Mat. Phys. \underline{14},
 21 (1988) and references therein.
\item{[7]} For chaotic behavior, see: A.J. Bray and M.A. Moore,
Phys. Rev. Lett. \underline{58}, 57 (1987); M. Nifle and H.J.
Hilhorst, Phys. Rev. Lett. \underline{68}, 2992 (1992).
\item{[8]} See, e.g. , K.H. Fisher and J. Hertz, {\it Spin Glasses
} (Cambridge University Press, Cambridge, 1991).
\item{[9]} S. Caracciolo, G. Parisi, S. Patarnello and N. Sourlas,
 Europhys. Lett. \underline{11}, 783 (1990); J. Phys. France
\underline{51}, 1877 (1990); J. Phys I \underline{1}, 627 (1991);
 D.A. Huse and D.S. Fisher,  J. Phys I\underline{1}, 621 (1991);
 E.R. Grannan and R.E. Hetzel, Phys. Rev. Lett. \underline{67},
 907 (1991). See also: E. Marinari et al, {\it Preprint}
cond-mat/9508036; C. Newman and D. Stein, {\it Preprint}
cond-mat/9508006.
\item{[10]} See, e.g. , D.P. Belanger and A.P. Young, J. Magn.
Magn. Mat. \underline{100}, 272 (1991) and references therein.
For SG phase see: F.C. Montenegro, M.D. Coutinho-Filho and S.M.
 Rezende, Europhys. Lett. \underline{8}, 382 (1989); S.M. Rezende
 et al, J. Phys. France \underline{49}, C8-1267 (1988). For
 crossover from RFIM to SG behavior see: F.C. Montenegro, U.A.
Leitao, M.D. Coutinho-Filho and S.M. Rezende, J. Appl. Phys.
 \underline{67} 5243 (1990); F.C. Montenegro et al., Phys. Rev.
 B\underline{44}, 2155 (1991-I); D.P. Belanger et al., ibid ,
 B\underline{44}, 2161 (1991-I).
\item{[11]} E.P. Raposo, M.D. Coutinho-Filho and F.C. Montenegro,
 Europhys. Lett. \underline{29}, 507 (1995).
\item{[12]} J. Mattsson, T. Jonsson, P. Nordblad, H. Aruga Katori
 and A. Ito, Phys. Rev. Lett. \underline{74}, 4305 (1995).
\item{[13]} C.M. Soukoulis, K. Levin and G.S. Grest, Phys. Rev.
 B\underline{28}, 1495 (1983); Phys. Rev. B\underline{33}, 7659 (1986).
\item{[14]} For weakly and moderatelly diluted systems see:
J.R.L. de Almeida and R. Bruinsma, Phys. Rev. B\underline{37},
 7267 (1987); U. Nowak and K.D. Usadel, Phys. Rev B\underline{44},
 7426 (1991-II); \underline{46} 8329 (1992-I).
%%%\end{References}
\end{itemize}

\newpage
\noindent
%%%\begin{Captions}
{\bf Figure Captions}\\\\
\begin{itemize}
\item[Fig.1.] (a) Part of a SG ``ground-state" spin configuration:
$x= 0.25$,  $T/T_{N}(x=1) = 0.18$ and $H = 0$; \underline{$\sqcap$}
 ($\sqcap $) indicate up (down) spins; $\circ $ indicate zero
local spins. (b) Local field distribution of the SG
``ground state".

\item[Fig.2.] Best fitting of the low $T$-dependence of the
correlation function: $G_{r}(T) = f(r)\exp (-r/\xi )$ , $\xi \sim
T^{(1/y_{T})}$, yelding $y_{T}= 0.19 \pm 0.03$ . The inset
shows the peak of $G_{r}(T)$ at the freezing temperature $T_{f}$.

\item[Fig.3.] (a) $T$-dependence of  $< S_{i} S_{j} >_{T}$ , with
 sites $i$ and $j$ randomly chosen on the lattice.
(b) Chaotic behavior
 of $< S_{i} S_{j} >_{T}$  as a function of spin distance at
$T/T_{N}( x = 1) = 0.18$ .

\item[Fig.4.] (a) Part of a FC spin configuration: $x= 0.25$,
 $T/T_{N}(x=1) = 0.18$ and $H = 1G$; $\sqcap $ indicate spin reversal
 relative to the SG ``ground state" of Fig. 1(a) . (b) Local
field distribution of a FC spin configuration. The inset shows a
ZFC local field distribution in the absence of frustration.

\item[Fig.5.] Effect of frustration on the $T$-dependence of
the EA order parameter for several values of fields
 (magnitude in Tesla): (a) FC cycle for $x = 0.25$; (b) ZFC
 cycle for $x = 0.48$ .

\item[Fig.6.] Field-induced AT irreversibility lines: dashed
lines indicate experimental estimate of the crossover exponent,
 $\phi \approx 3.4$; dot-dashed lines indicate the AT value,
$\phi = 3.0$; full lines are best-fitings. The insets show the
 $T$-dependence of the correlation function: $\bullet $
indicate data at $H=0$; $\sqcap $ ($\diamond $) are FC (ZFC)
 data at $H=0.5$T; (a) in the absence of frustation,
 $\phi = 3.6 \pm 0.9 $ ( notice that $G_{r}(T)$ is null at $H=0$ );
 (b) in the presence of frustration, $\phi = 3.8 \pm 0.6$.

%%%\end{Captions}
\end{itemize}

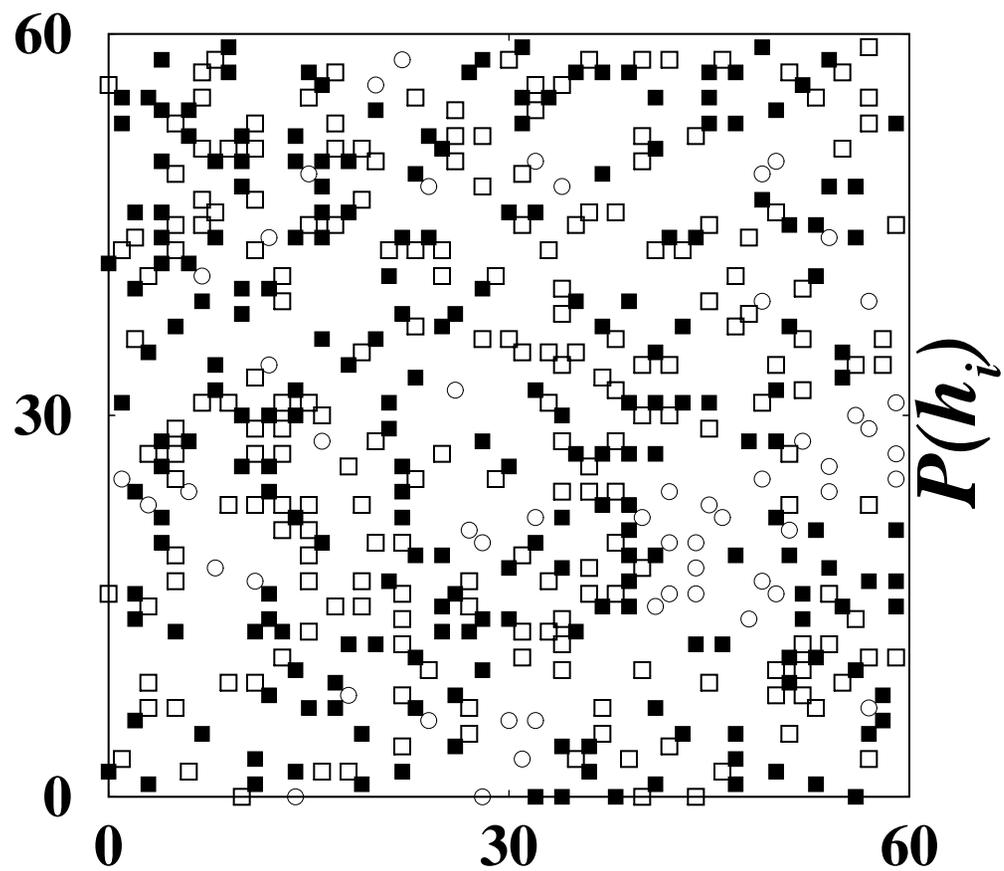
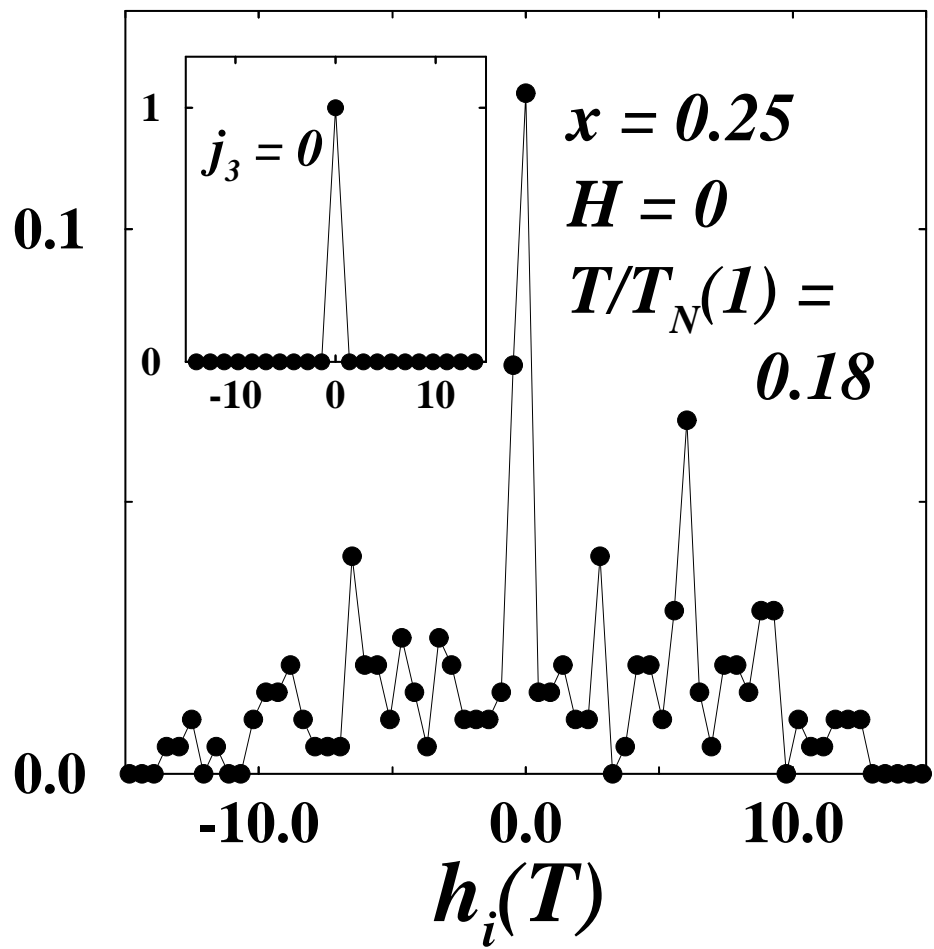
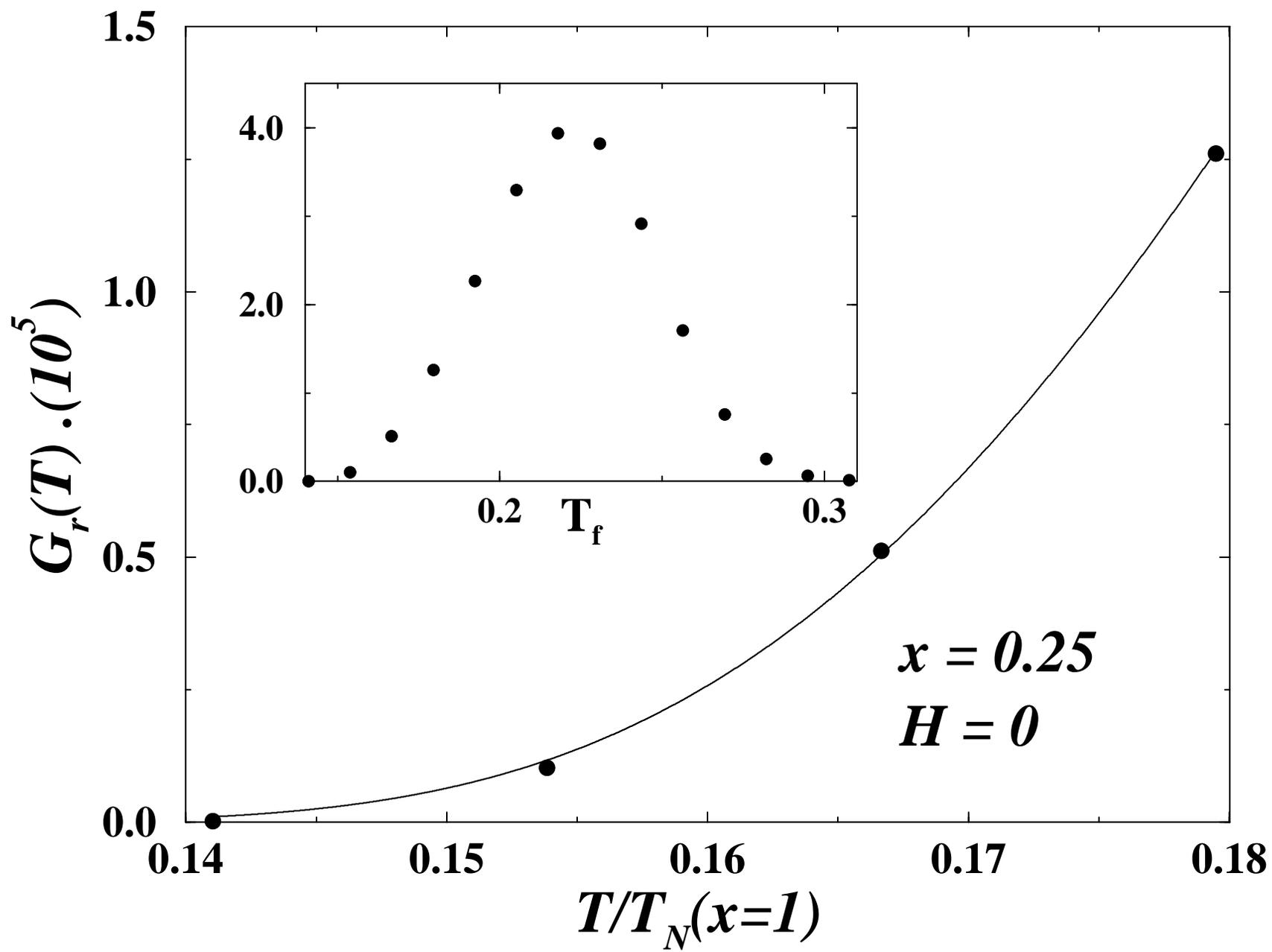
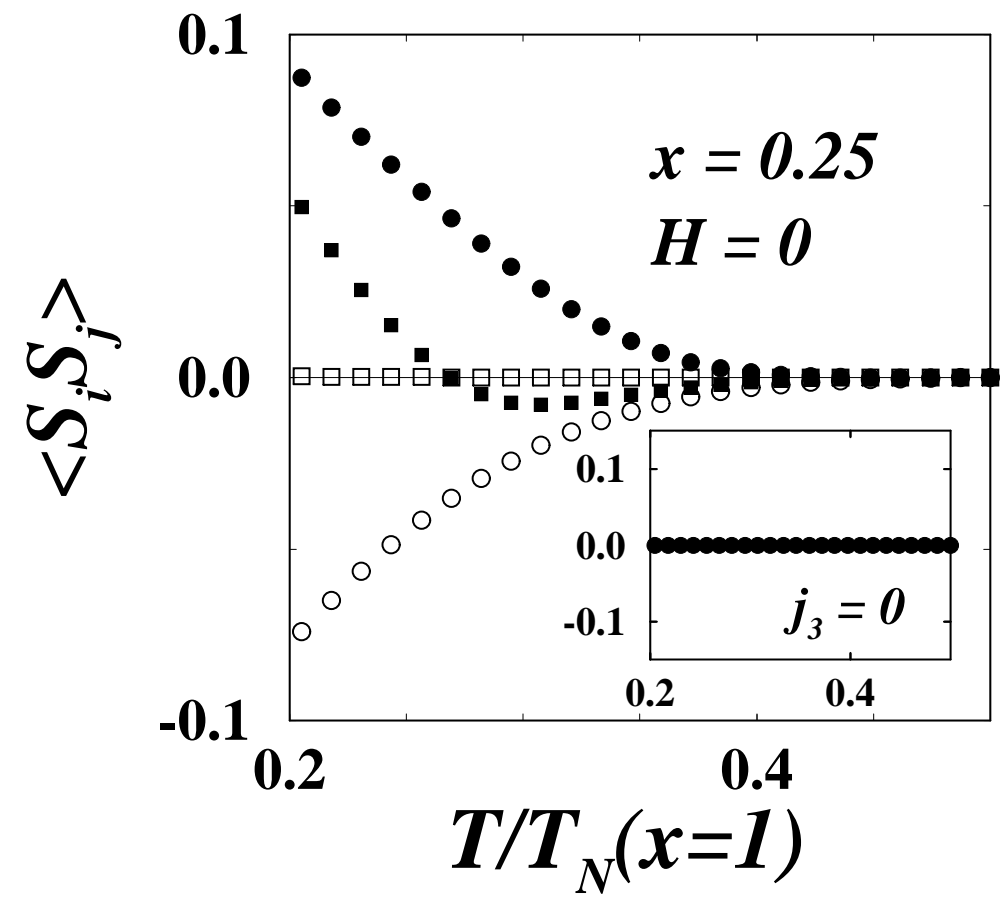
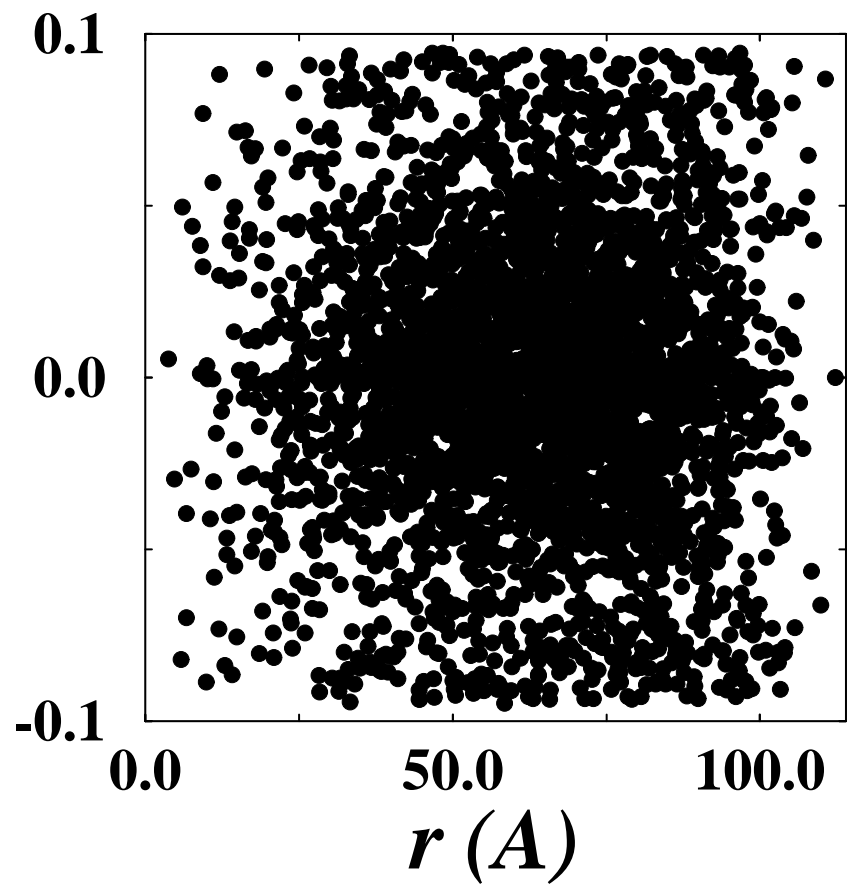
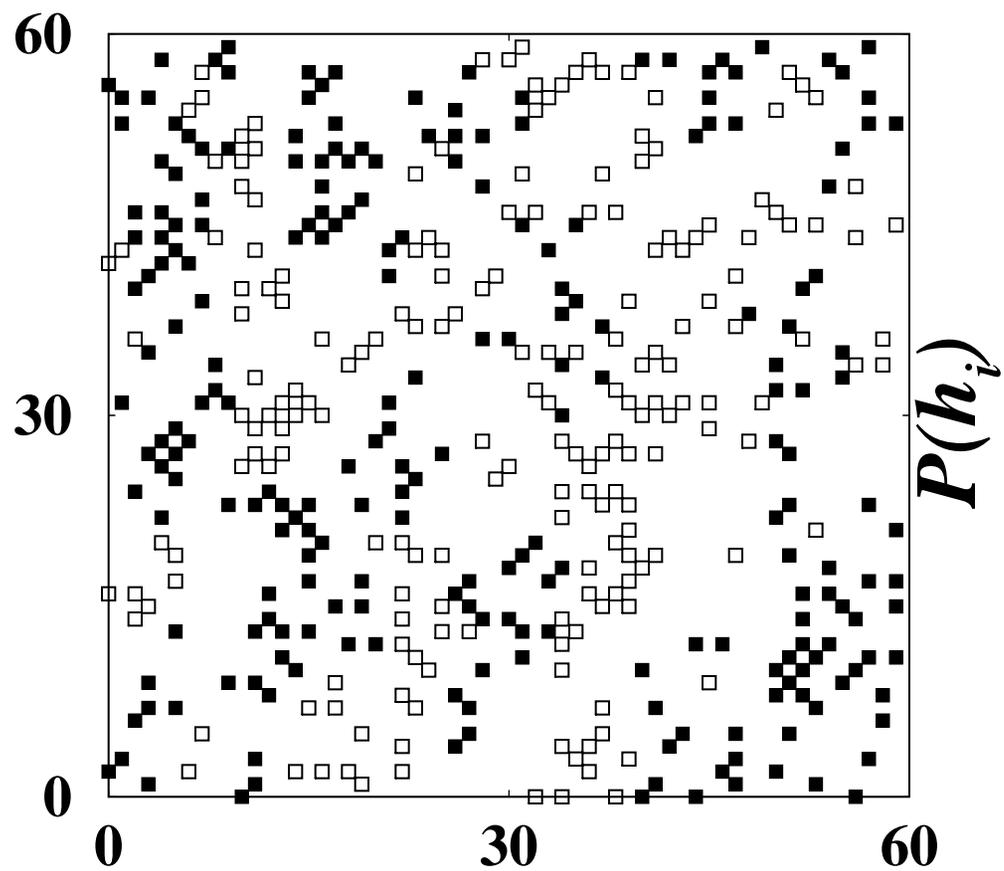
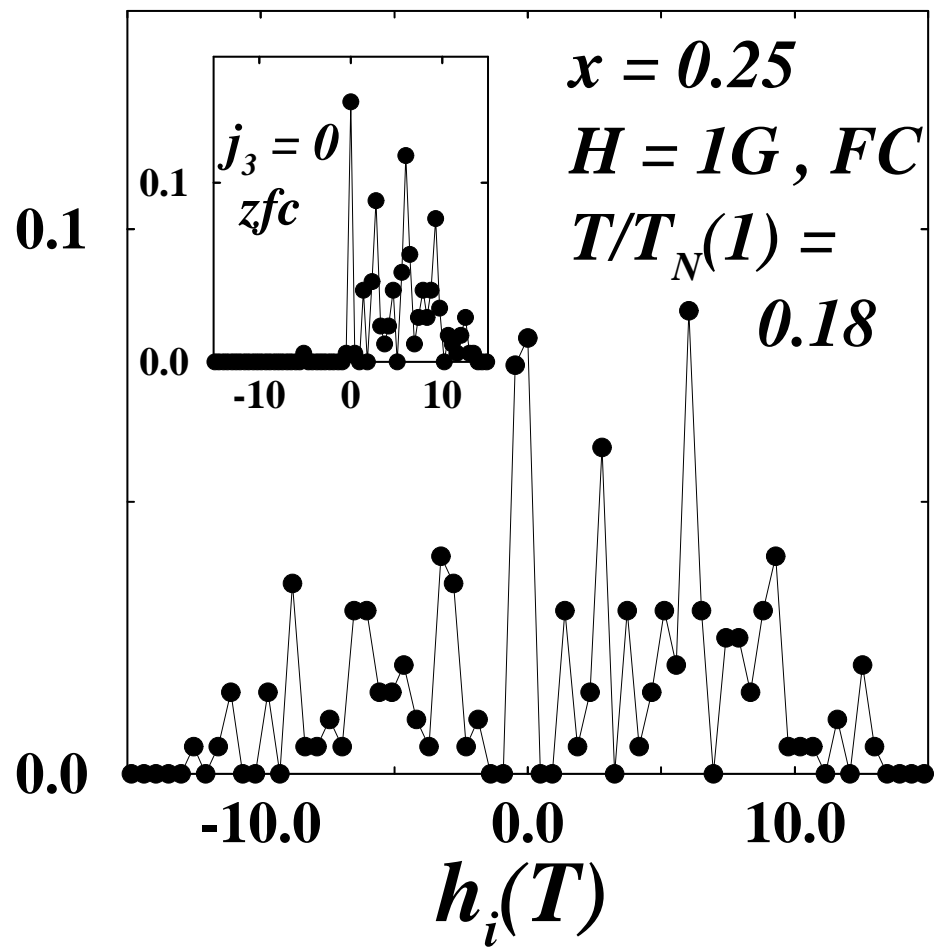
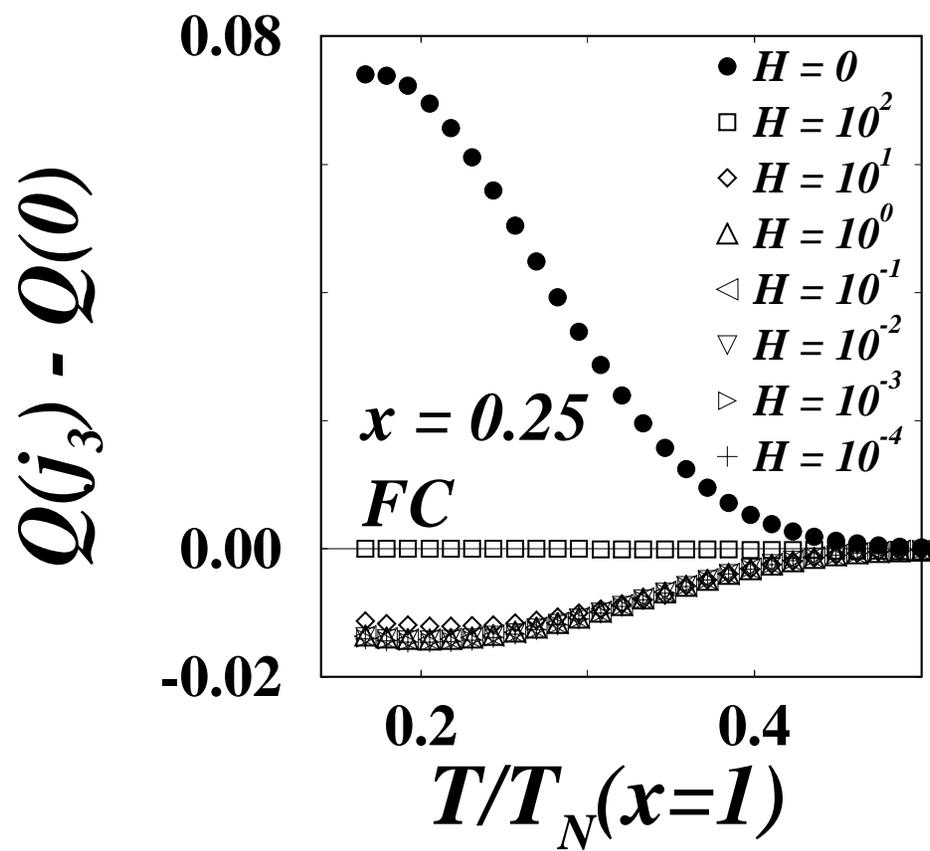
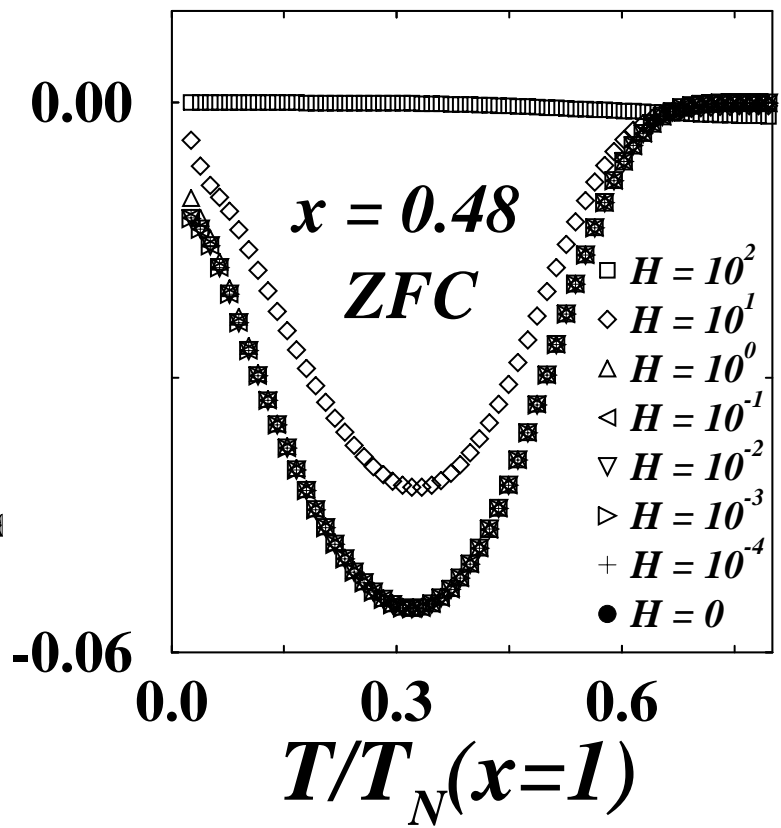
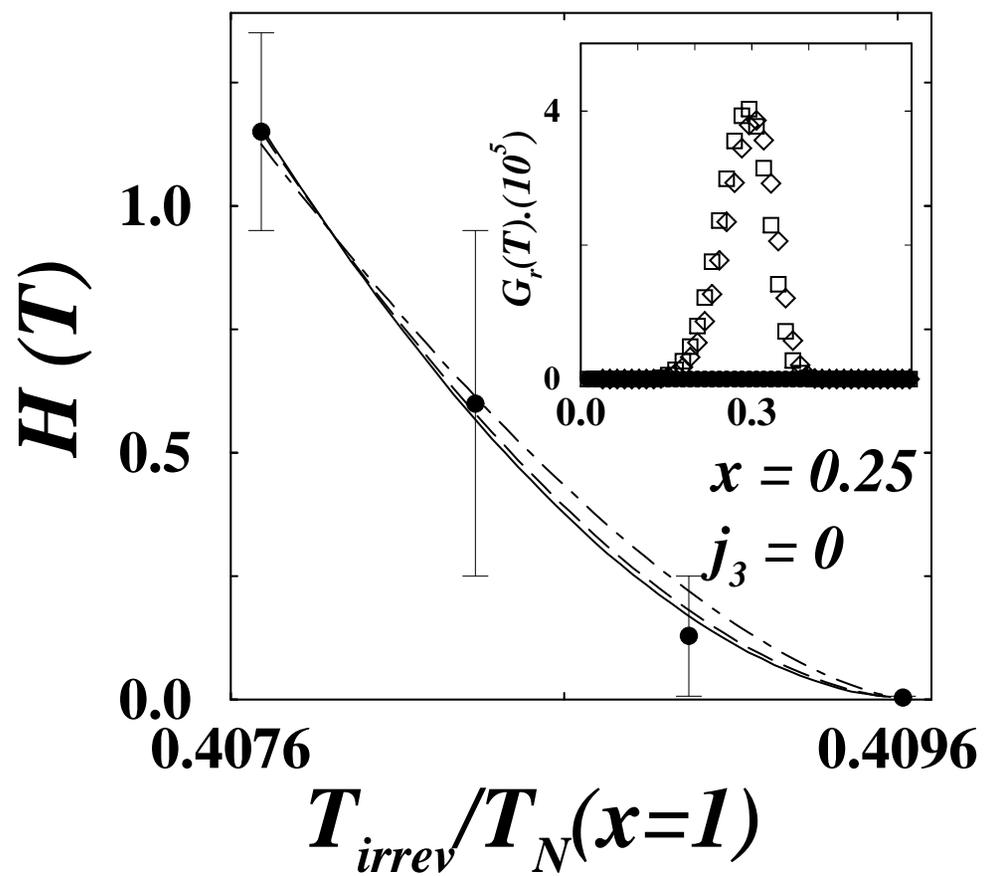
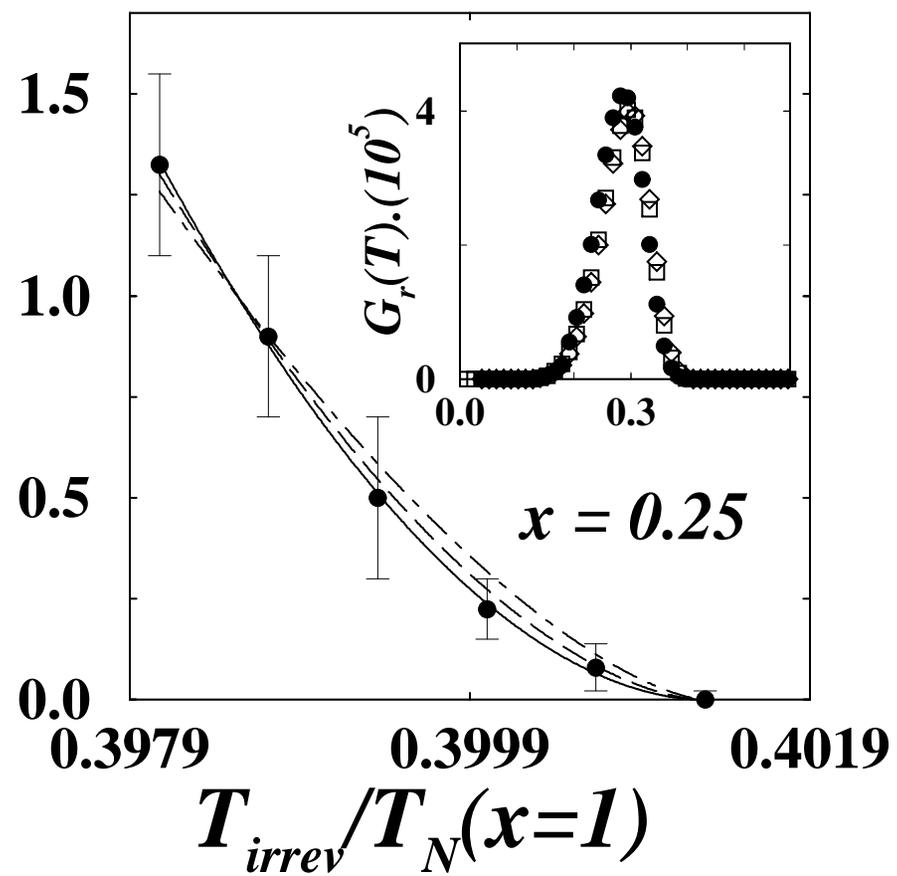
\end{document}